\documentclass[preprint,prc,aps,showpacs,superscriptaddress]{revtex4}
\usepackage{epsfig}
\usepackage{amsmath}
\usepackage[hyperref]{hyperref}
\usepackage{epstopdf}
\begin{document}

\preprint{KEK-TH-1328}

\title{Using  branching processes in nuclei  to reveal dynamics of large-angle, two-body scattering }

\author{S. Kumano}
\email[E-mail: ]{shunzo.kumano@kek.jp}
\affiliation{KEK Theory Center, Institute of Particle and Nuclear Studies \\
          High Energy Accelerator Research Organization (KEK) \\
          1-1, Ooho, Tsukuba, Ibaraki, 305-0801, Japan}
\affiliation{Department of Particle and Nuclear Studies \\
             Graduate University for Advanced Studies \\
           1-1, Ooho, Tsukuba, Ibaraki, 305-0801, Japan}     
\author{M. Strikman}
\email[E-mail: ]{strikman@phys.psu.edu}
\affiliation{Department of Physics, Pennsylvania State University \\
             University Park, PA 16802, U.S.A.} 

\begin{abstract}
We demonstrate  that hard branching $2\to 3$  particle  processes with nuclei provide 
an effective way to determine the  momentum transfers needed for effects of point-like  configurations  to  dominate  large angle $2\to 2$ processes. 
In contrast with  previously proposed approaches, the discussed
reaction allows  the effects of   the transverse size of
configurations to be decoupled from effects of the space-time
evolution of these configurations.  
\end{abstract}

\pacs{13.85.-t, 13.60.Le, 12.38.-t}  

\maketitle

%%%%%%%%%%%%%%%%%%%%%%%%%%%%%%%%%%%%%%%%%%%%%%%%%%%%%%%%%%%%%%%%%%%%%%%%%%%%%%%%

Large angle two body processes seem to be amongst the simplest hadronic interactions 
 that can  be described by perturbative quantum
chromodynamics (pQCD). In the limit of $s \to \infty, t/s=const$ these processes are dominated by  hard gluon exchanges between the constituents with all of the  quark constituents at small relative distances \cite{bf7375,Lepage:1980fj}.
Contributions of  large size configurations (in particular the end point contribution) should be suppressed in this limit by  Sudakov form factors, see discussions  in 
Refs.~\cite{Mueller:1981sg,Botts:1989kf}.
 Nevertheless, after many years and many  studies the momentum
transfer range of applicability of pQCD has not been firmly established. A series of experiments at Brookhaven National Laboratory (BNL)   measured the nuclear transparency of nuclei in 
quasielastic scattering process near 90$^\circ$  in the pp center of mass, see the  summary in \cite{Aclander:2004zm}.
 An   increase of nuclear transparency
was observed  between $p_N^{inc} =\mbox{5.9  GeV/c}$
and $p_N^{inc} = \ \mbox{9.5 GeV/c}$,  indicating  that freezing of nucleonic configurations  becomes possible for  $p_N \ge \ \mbox{8 GeV/c}$. This rise is  followed by  drop of the transparency at larger incident momenta indicating  that 
some nonperturbative mechanisms play an important role 
in  nucleon-nucleon scattering (for the reviews see e.g.
\cite{Miller:2007zzd,Jain:1995dd}) up to the momentum transfer squared
$-t \sim 13 \ \mbox{GeV}^2$. 
At the same time,  the observed enhancement of the $K^+p $ elastic cross section as compared to $\pi^+ p$ elastic cross section 
for $\theta_{c.m.}=90^0$ and $-t \ge 5 \ \mbox{GeV}^2$ suggests that scattering processes are dominated by  point-like configurations (PLC)  in mesons that  have a larger probability for the  $K^+$-meson  than for the pion 
(see discussions in the beginning of Sec.II and also in the end of
Sec.VI of Ref. \cite{Kumano:2009he}). In principle, the onset of the regime
of dominance of PLC could be quite different for meson and baryon projectiles because 
 a meson is a much  simpler object than a baryon. A larger  PLC probability is obtained for the simple reason that  only two quarks have a  close encounter. In addition, the  nonperturbative structure of baryons is likely to be  much more complicated. This is indicated in particular by the structure of hadrons  in the large $N_c$ limit in which meson remains a $q\bar q$ system
while a nucleon can be viewed as a soliton \cite{Witten:1979kh}. 
 It was nearly three decades ago  when Refs.~\cite{Mueller82,Brodsky82} suggested
 testing  the mechanism of these reactions using the color cancellation (CC) property  of color-neutral objects
 of 
QCD - the suppression of the interaction of small size
color singlet wave packets with hadrons. CC plays a key role in
ensuring approximate Bjorken scaling in deep inelastic scattering
\cite{fms94}, in proving QCD factorization theorems for high energy
hard exclusive processes \cite{factorization}, etc.    It also leads to color transparency CT 
under certain conditions (see discussion below).
CC may  be visualized in  the high energy limit by  introducing 
a notion of the scattering cross section of a small dipole 
configuration (say $q\bar q$) with transverse size $d$ on the nucleon
\cite{Frankfurt:1993it, Blaettel:1993rd} which in the leading log
approximation is given by Refs. \cite{Frankfurt:2000jm,BP-book}
\begin{equation}
\sigma(d,x)= {\pi^2\over 3} \alpha_s(Q^2_{eff}) d^2
   \left[x G_N(x,Q^2_{eff}) + \frac{2}{3} x S_N(x, Q^2_{eff})\right],  
\label{pdip}
\end{equation}
where $Q^2_{eff} = \lambda/d^2, \lambda= 4 \div 10$ \cite{lambda},
$G_N$ is the nucleon gluon distribution, $S_N$ is the  sea quark distribution 
for quarks making up the dipole, $x$ is the momentum fraction carried
by a parton, and $\alpha_s$ is the running strong-interaction coupling constant.    
Note that Eq.(\ref{pdip}) predicts a  substantially more rapid
increase of the dipole-hadron cross section with increase of energy 
$\propto xG_N(x,Q^2_{eff}) \propto x^{-n(Q^2_{eff})},    
n(4 \ {\rm GeV}^2) \sim 0.2$ than for the soft processes.
This expectation is qualitatively different from the expectation
of the two gluon exchange model where the  cross section does not
depend on energy \cite{Low:1975sv}. CT for high energy scattering
from nuclear targets  was observed for coherent $J/\psi$ production \cite{Sokoloff:1986bu}.
The  experiment \cite{Aitala:2000hc} performed at Fermi Lab with 
the 500 GeV pion beam  confirmed  the  key CT predictions
of Ref.\cite{Frankfurt:1993it} . In particular, the authors reported  a 
 strong increase of
the cross section in  the $\pi +A \to ``{\rm two \ jets}" + A$   
process with $A$ ($A$=carbon and platinum):     
$\sigma \propto A^{1.61\pm 0.08}$ as compared to the prediction
$\sigma \propto  A^{1.54}$.

The prediction of increase of transparency with exclusive light 
vector meson production \cite{Frankfurt:1988nt,Brodsky:1988xz}
 is consistent with indictions of the FNAL E-665 \cite{Adams:1994bw}
 and HERMES \cite{Airapetian:2002eh} data on the 
 $\rho^0$ leptoproduction (though these data were taken in the kinematics which did not exclude production of hadrons in the nucleus fragmentation region).

At intermediate energies observing  CT (in the kinematics where CC holds) is complicated by the effects of quantum diffusion \cite{FLFS88,fms94,jm}.   
Even if PLCs of hadrons are involved in the collisions
the space-time evolution leads to expansion of the wave packets as they move away from the interaction point, so that at a distance, $l_{coh}$
which is referred to as the coherence length 
\begin{equation}
l_{coh} \sim l_0 \ \mbox{fm} \cdot  p_h/\mbox{GeV}, 
\label{lcoh}
\end{equation}
the packet expands to a normal hadronic size. Theoretical estimates
give $l_0$ in the range $l_0 \sim (0.35 \div 0.8) \ \mbox{fm}$,
see a review in Ref.\cite{fms94}. Here $p_h$ is the hadron momentum
in GeV. This leads to a strong reduction of the CT effect over
a wide range of incident energies. In particular the estimates
of Ref.\cite{FLFS88} indicate that the effect of CT in say 
$A(p,2p) $ reactions at $\theta_{c.m.} \sim 90^\circ$ is greatly
reduced over  a wide range of energies $\le $ 20 GeV.   
A high resolution experiment of pion production recently reported 
evidence for the onset of CT \cite{:2007gqa} in the process
$eA\to e\pi^+ A^*$.  These experimental results  agree well with
predictions of Ref.\cite{Larson:2006ge} where the effects of  CT were calculated using  the quantum diffusion model with  Eq.(\ref{lcoh})  and a value of 
 $l_0 = 0.57 \ \mbox{fm}$. 
This confirms the  small scale  of $l_{coh}$ (a larger $l_{coh}$ would
lead to a stronger CT effect for heavy nuclei) and indicates
that it would be very difficult to determine the degree of
squeezing of the hadronic configurations in the $2\to 2$
processes at a wide range of momentum transfers using
the $A(h, h'N)$  reactions. The difficulty is that understanding the dynamics are forces  us 
to  study two phenomena at the same time - squeezing
at the initial point and the pattern of expansion.

%%%%%%%%%%%%%%%%%%%%%%%%% figure %%%%%%%%%%%%%%%%%%%%%%%%%
\begin{figure}[b]
\begin{center}
\includegraphics[width=0.40\textwidth]{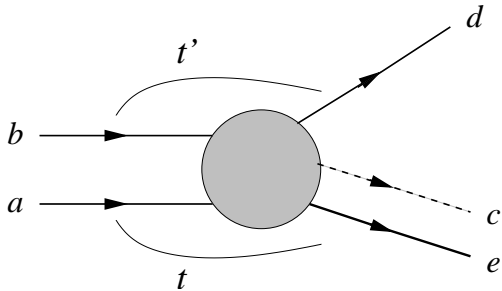}%{diag1.pdf}
\vspace{-0.3cm}
\caption{$a+b \rightarrow c + d + e$ reaction.}
\label{fig:diagram1}
\end{center}
\end{figure}
%%%%%%%%%%%%%%%%%%%%%%%%% figure %%%%%%%%%%%%%%%%%%%%%%%%%

\begin{figure}
\includegraphics[width=0.8\columnwidth]{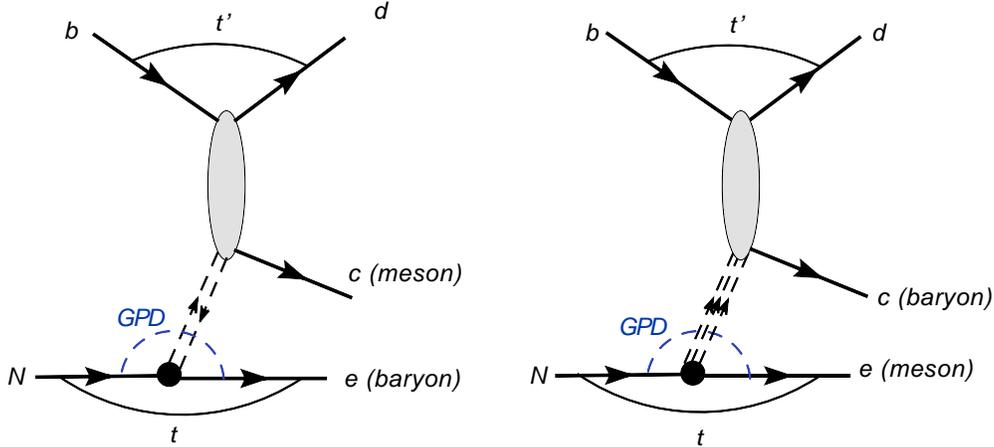}
\vspace{-0.2cm}
\caption{$(a)$ Production of fast meson and recoiling baryonic system. 
         $(b)$ Production of fast baryon and recoiling mesonic system.}
\label{diagram:Fig2}
\end{figure}

Here we suggest a new strategy which allows the suppression of  effects
related to the space-time evolution of the wave packet 
and also allows  checking  the onset of  
the dominance of the contribution of PLC at moderate energies.
This strategy uses  novel hard branching $2\to 3$  processes \cite{Kumano:2009he}
(see Fig. \ref{fig:diagram1})
\begin{equation}
a+b \to c + d + e.
\end{equation}
The reaction is considered in the limit \cite{Kumano:2009he}:
\begin{equation}
- t'= - (p_b-p_d)^2 \to \infty, \ \ s'=(p_c+p_d)^2 \to \infty,        
             \ \ \mbox{and} \ -t'/s' \to {\rm const},  
\end{equation}  
and  
\begin{equation}
- t=-(p_a-p_e)^2 = {\rm const} \leq m_N^2,  
\end{equation}   
where $p_i$ is the momentum of hadron $i$ 
($i=a,\ b,\ \cdot\cdot\cdot,\ e$) and $m_N$ is the nucleon mass.               
In this limit the leading-order QCD diagrams dominate
the cross section for two-body processes \cite{bf7375}.
As a result one can provide formal arguments \cite{Kumano:2009he}
that the amplitude of the process is factorized into a product
of the generalized parton distribution and the amplitude of
large angle scattering of the projectile ``$b$"    
off point-like $q\bar q$ or $3q$ configuration, see Fig.\ref{diagram:Fig2}.

%%%%%%%%%%%%%%%%%%%%%%%%%%%%%%%% figure %%%%%%%%%%%%%%%%%%%%%%%%%%%%%%%%
%%%%%%%%%%%%%%%%%%%%%%%%%%%%%%%% figure %%%%%%%%%%%%%%%%%%%%%%%%%%%%%%%%

In the case when $b$ is the projectile,  hadrons $d$ and $c$ carry
practically all momentum  of $b$, while the recoil particle carries
small energy in the lab. frame:
\begin{equation}
\vec{p}_d= (xp_b, p_t^{d}), \ \       
\vec{p}_c= ((1-x)p_b, p_t^{c}), \ \   
p_t^{d}\approx - p_t^{c} \equiv p_t,  
\end{equation}
and
\begin{equation} 
s' \simeq  \frac{m_d^2 + p_t^2}{x} + \frac{m_c^2 + p_t^2}{1-x}, \ \    
-t \simeq  \frac{(s' -m_b^2)^2}{s} .    
\end{equation}

In kinematics with $x\sim 1/2$, two leading particles carry large
longitudinal momenta of about one half of the projectile momentum. 
This feature of the discussed process gives it a great advantage
for investigating  the dynamics of the large angle $2\to 2$ processes
using nuclear targets. Indeed, in contrast with the elementary
$2\to 2$ process,  in the case of the $2\to 2$ process embedded
in the $2\to 3$ process there is no correlation 
between momentum of the projectile and the value of $p_t$
of the produced hadrons $d,c$.
 In the $2\to 2$ process at a fixed scattering
angle ($e.g.$ $\theta_{c.m.}=90^\circ$), there is one to one
correspondence between the projectile momentum and the value of $p_t$.
In a sense, using  the $2\to 3$ kinematics lets one  boost hadrons
in the PLCs relative to the nucleus, thereby  practically completely 
freezing the PLCs. This is achieved in the limit when we  keep  $p_t$ and mass of the $c$, $d$ pair the same, hence preserving 
 the kinematics of hard scattering but allowing 
 the total momentum
of the pair to vary. 
As a result,  one can regulate the degree of 
 freezing of
the pair while it propagates. 

As an example, let us consider a  test of  CT in elastic
$\pi^+ \pi^- $ scattering. We choose this example for several reasons:
Firstly, due to the minimal number of constituents in this process 
we expect a lower-energy onset of the CT regime than, say, in $pp$ scattering.
Secondly, the rate of the space-time evolution in this case is 
constrained by the  pion electroproduction data \cite{:2007gqa}.
 Thirdly, it is known 
that  the  cross section of the elementary process 
\begin{equation}
\pi^- p \to \pi^- \pi^+ n,
\end{equation}
is sufficiently large as it was measured at FNAL at 100 GeV/c 
and 175 GeV/c including kinematics where $s', -t' $ are of the order of
few GeV$^2$ \cite{Bromberg:1983kv}. The data indicate  dominance of
the pion exchange in $t$-channel which is consistent with the expectations
for the hard kinematics as the pion-pole dominated GPDs
(generalized parton distributions) give significantly larger contribution
than the $\rho$-pole dominated GPDs, cf. analysis of the 
 $2\to 3$ \, process $NN\to N\pi N$ in Ref.\cite{Kumano:2009he}. 
Also, the COMPASS collaboration at CERN has collected large statistics
for forward pion production for 190 GeV pion scattering off
a wide range of nuclei \cite{COMPASS} and has plans for further
measurements using the pion beam. There are other interesting channels
for measurements with pion and proton beams. We  briefly discuss these channels at the end of this article.

Since coherent scattering  for a nuclear
target, is negligible for this reaction,  the process we examine  is $\pi^- A \to \pi^+ \pi^- A^*$,
where the total energy of the residual system is close to $M_A -t/2m_N$. 
The underlying elementary process involves transition of proton to
neutron, and it generates the final system
of $A$ nucleons with a small overlap with  a nuclear-bound state.
In principle, it may be difficult to experimentally exclude  the production
of an  excited 
 hadronic system, like the  $\Delta$-isobar 
(for example $\pi^- n \to \pi^-\pi^+ \Delta^-$). 
However the factorization theorem holds for any fixed mass of
the produced system ``$e$". Typically, the experimental momentum resolution 
\cite{COMPASS} for the detected pions is the same for different
nuclei/hydrogen ($^2$H) targets. In this case, the cuts on the mass
of the produced hadronic system ($N, \Delta, N^*,...$) remain the same
and would not affect our predictions for transparency.

First we consider CT effects for  high  energy projectiles
- $E_{\pi} \sim 200 \ \mbox{GeV}$. In this case, Eq.(\ref{lcoh}) tells us 
that the coherence length of the final pions exceeds 30 fm $\gg R_A$,
and is a factor of two larger for the projectile pion. Therefore 
expansion effects may be neglected. We define nuclear transparency as 
\begin{equation}
T_A= {{d \sigma (\pi^- A \to \pi^-\pi^+ A^*) \over d\Omega}\over 
Z {d \sigma (\pi^- p \to \pi^-\pi^+ n) \over d\Omega}},
\end{equation}
where $\Omega$ is the solid angle for the $\pi^-\pi^+$ system.
This  ratio can be estimated using the semi-classical approximation as \cite{FLFS88} 
\begin{equation}
T_A(\vec p_b,\vec p_c,\vec p_d) = {1\over A} \int d^3 r \rho_A(\vec r)  
P_b (\vec p_b, \vec r) P_c (\vec p_c, \vec r) P_d (\vec p_d, \vec r),    
\label{TA}
\end{equation}
where $\vec p_b,\vec p_c,\vec p_d$ are three momenta of the incoming
and outgoing particles $b$, $c$, $d$; $\rho_A$ is the nuclear density
normalized to $\int \rho_A(\vec r) d^3 r =A$ (for simplicity we neglect 
here a small difference between  the proton and matter distributions).
The probabilities of  no interaction with the entering and
2 outgoing fast hadrons are given by the product of probabilities $P_j$  
\begin{equation}
P_j (\vec p_j,\vec r)=\exp\left(-\int_{\rm path} d z \,                       
                \sigma_{\rm eff} (\vec p_j, z) \rho_A(z)\right).   
\label{psigma}
\end{equation}
Here $P_j$ is the   probability for particle $j$ with momentum $\vec p_j$ to
propagate along a path from the point of hard interaction $\vec r$
and $z$ is the distance from the interaction point. Here, for generality, we write the expression allowing for expansion effects.

For  soft interactions the effective cross section is given by   
$\sigma_{eff} \sim \sigma_{tot}(\pi N)$. As a result, in this limit
Eqs.(\ref{TA}),(\ref{psigma}) predicts values of  transparency that drop strongly with increasing values of $A$, and which
are dominated by the scattering off nucleons of the rim of the 
nucleus  $\propto  A^{1/3}$ . For example, for the case of a soft interaction ($\sigma_{eff}=25$ mb)
$n(A) =  \partial \ln \left(AT(A)\right)/\partial \ln A$ is about $  0.30 \ (0.24)$
for A= 40 (208).
In the high energy limit,
when the $q\bar q$ configurations of incoming and outgoing pions
can be considered as frozen $T(A)$ can be estimated using Eqs.(\ref{TA}),(\ref{psigma}) with $z$-independent $\sigma_{eff}$.
For the purpose of obtaining rough estimates we will neglect  a possible difference
in the  degree of squeezing of incident and outgoing pions as well as
the energy dependence of the cross section as given by Eq.(\ref{pdip})
(the second effect is definitely small as the gluon density changes
in the discussed virtuality range less rapidly  
than $x^{0.2}$ which
corresponds to the difference of the cross sections for initial
and final pions of $2^{0.2} \sim 1.15$).  We find that $T_A$ is
very sensitive to a variation of $\sigma_{eff}$ - see Fig.3.
%%%%%%%%%%%%%%%%%%%%%%%%% figure %%%%%%%%%%%%%%%%%%%%%%%%%
\begin{figure}
\includegraphics[width=0.8\columnwidth]{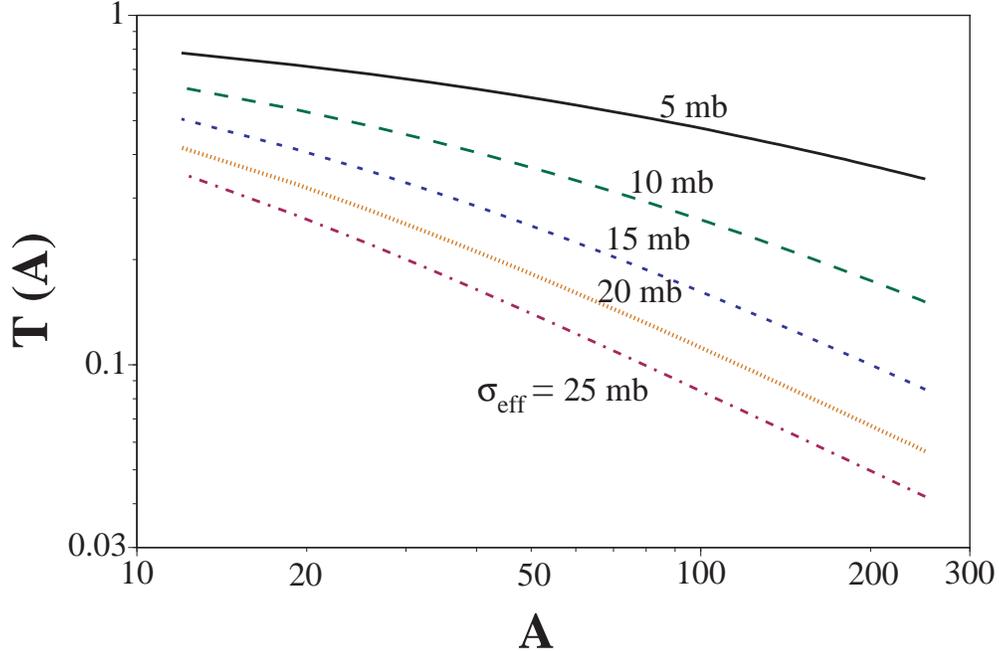}
\vspace{-0.2cm}
\caption{$A$-dependence of $T(A)$ for different values 
         of $\sigma_{\rm eff}$. } 
\label{diagram:Fig3}
\end{figure}
%%%%%%%%%%%%%%%%%%%%%%%%% figure %%%%%%%%%%%%%%%%%%%%%%%%%
The degree of squeezing can be estimated by  considering  
 the leading QCD diagrams for $\pi$-$\pi$ scattering.
For a pion with a given $p_t$, the  internal characteristic momenta are 
$\sim p_t/2$, corresponding to the transverse size $d$ of the dipole
of the order ${\pi/2\over p_t/2}$. For $p_t= 1.5\, (2) \,{\rm GeV}/c$
this corresponds to $d= 0.4\, (0.3)\, {\rm fm}$ where Eq.(\ref{pdip})
for a $q\bar q$ wave packet  with energy 100 GeV gives 
$\sigma_{eff} \le 4 \, {\rm mb}$. 
Hence, if the perturbative mechanism of  $\pi$-$\pi$ scattering
dominates in this $p_t$ range, a large color transparency effect is
predicted. For example, one can see from Fig.3 that  for $A=40 \ (208)$, 
transparency $T(A)$ is predicted to increase by a factor $\sim 4 \ (8)$ 
from its ``Glauber-type"   
geometric value of 
$\sigma_{eff}=25 \,\mbox{mb}$ to $\sigma_{eff}=5 \,\mbox{mb}$.
Since for small $p_t$ one expects the geometric picture with 
$\sigma\sim \sigma_{\pi N}$ to describe $T(A)$ reasonably well, 
we expect that the ratio $T_{A_1}/T_{A_2}$ should strongly depend
on $p_t$. In the regime of small absorption one can determine
$\sigma_{eff}$ from a  study of A-dependence and, by  using Eq. (\ref{pdip}),
determine the average size of the color dipoles involved in the process.

%%%%%%%%%%%%%%%%%%%%%%%%%%%%%%%%%%%%%%%%%%%%%%%%%%%%%%%%%%%%%%%%%%%%%%%%%%%%%%%%

{\it Study of the space-time evolution of small wave packets}

\begin{figure}
\begin{tabular}{cc}
\includegraphics[width=.45\textwidth]{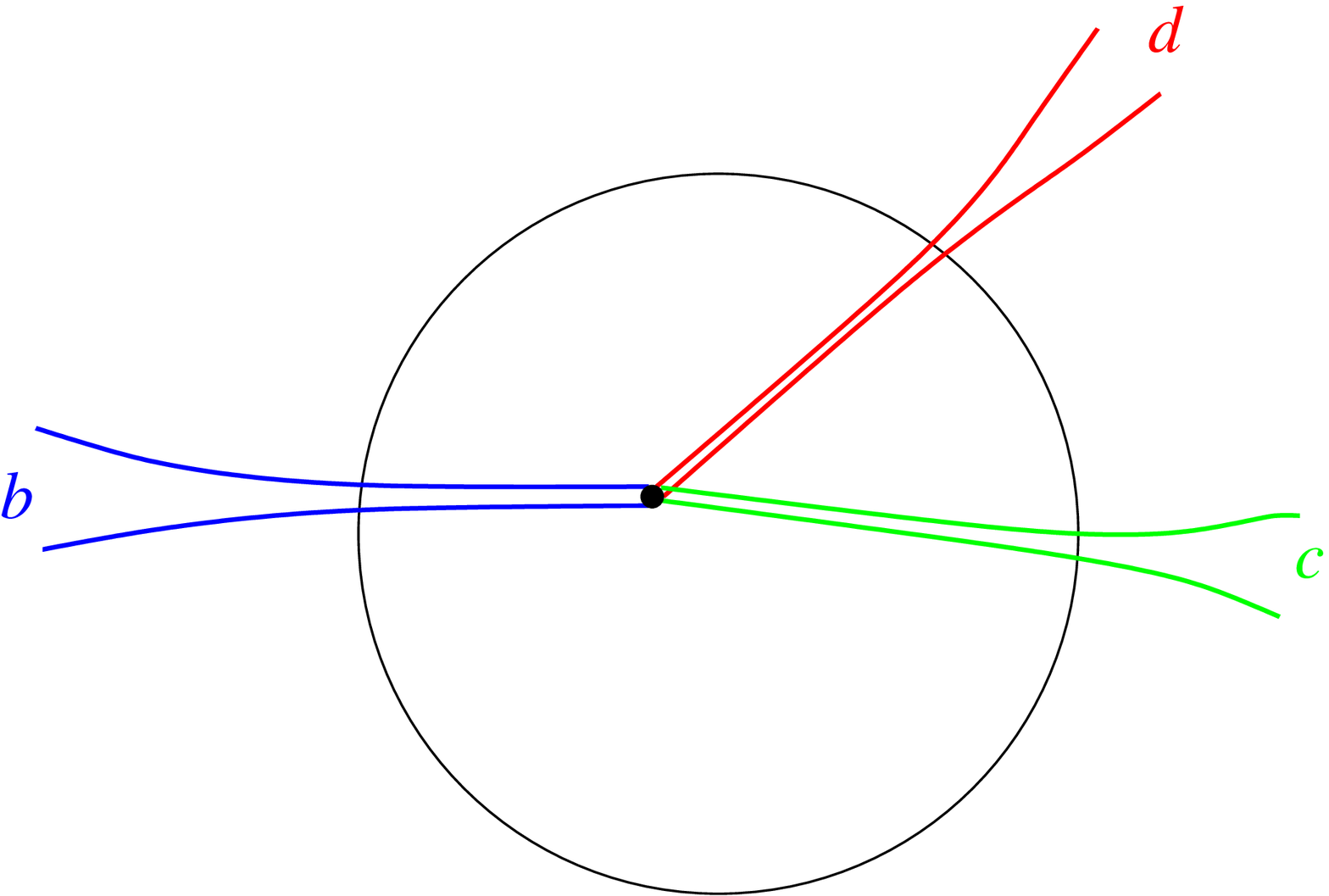}
&
\includegraphics[width=.45\textwidth]{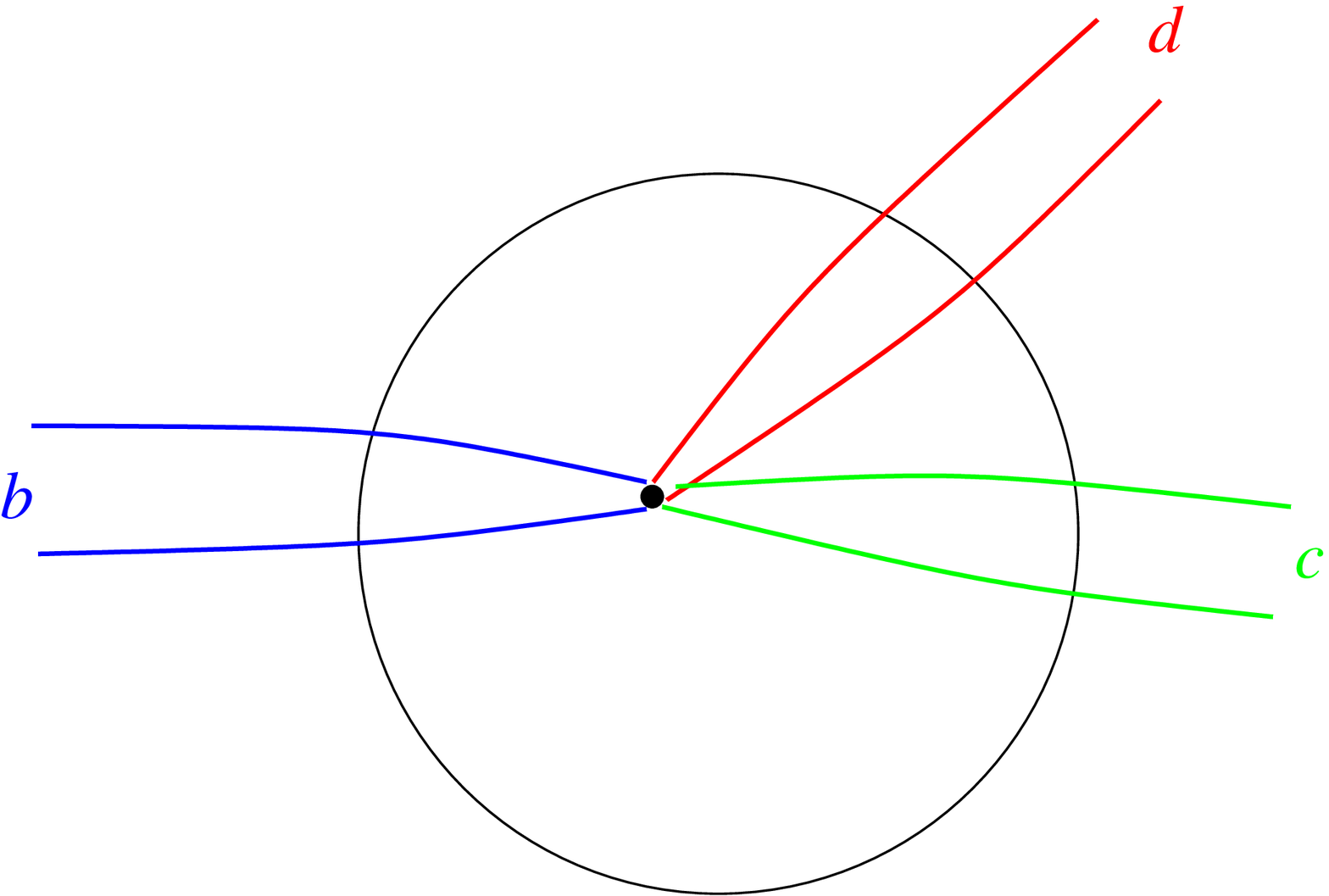}
\\[1ex]
(a) & (b) 
\end{tabular}
\caption[]{Space-time evolution of the wave packets of incoming and outgoing fast hadrons for large (a)  and  moderate incident momenta (b).} 
\label{expansion}
\end{figure}

Assuming that the measurements at large energies observe the effects of CT, 
a next step would be to study $T(A)$ for production of the $\pi \pi$ pair 
for fixed $s',t'$ as a function of the incident momentum, $p_{\pi}$. Indeed in this limit the $2\to 3$ amplitude is factorized into the product of the amplitude describing the  hard block  of $2\to 2$ process and the GPD describing coupling  to the soft block. Therefore,   the sizes of the hadronic configurations  involved in the $2\to 2$ large $s',t'$ process should not depend on $p_{\pi}$ at  the interaction point. Hence the $p_{\pi}$ dependence of $T(A)$   under these conditions should originate from the contraction of the small size configuration in the projectile $b$ as it approaches the interaction point and expansion of the outgoing wave packages which evolve into hadrons $c$ and $d$. At large values $p_{\pi}$ contraction and  expansion occur outside the nucleus (Fig.\ref{expansion}a), while with 
decreasing values  of $p_{\pi}$ both contraction and evolution  occur inside the nucleus (Fig. \ref{expansion}b).
To estimate the expected effect we can use the quantum diffusion model
of Ref.\cite{FLFS88} which gives
\begin{equation}
\sigma_{eff}(z)=  \left(\sigma_{hard}+{z\over l_{coh}}
\left[\sigma_{soft}-\sigma_{hard}\right]
\right) \theta(l_{coh}-z) + \sigma_{soft}\theta(z-l_{coh}),
\end{equation}
where $z$ is the distance from the interaction point, $l_{coh}$ is
given by Eq.(\ref{lcoh}), $\sigma_{hard}$ is the interaction of the PLC 
close to the interaction point and the interaction reaches the strength of 
soft interaction at $z=l_{coh}$, so $\sigma_{soft} \sim \sigma_{tot}(\pi N)$.
%%%%% 
%%%%%%%%%%%%%%%%%%%%%%%%% figure %%%%%%%%%%%%%%%%%%%%%%%%%
\begin{figure}
\includegraphics[width=0.8\columnwidth]{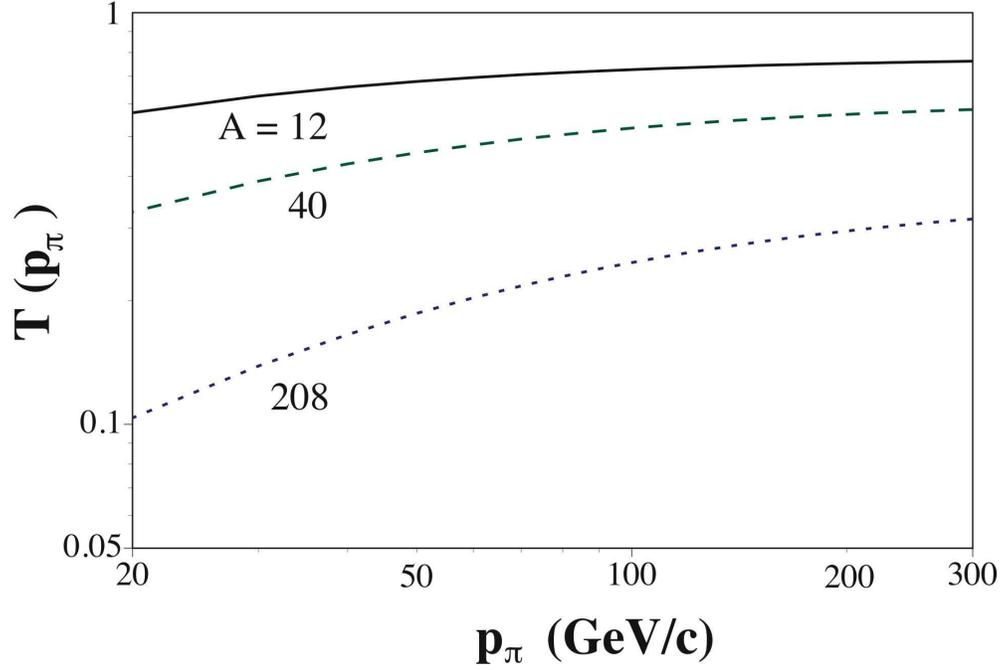}
\vspace{-0.2cm}
\caption{$p_\pi$-dependence of $T(A)$ for different nuclei
and $\sigma_{hard}= 5 \ \mbox{mb}$. }  
\label{diagram:Fig4}
\end{figure}
%%%%%%%%%%%%%%%%%%%%%%%%% figure %%%%%%%%%%%%%%%%%%%%%%%%%

We performed numerical calculations using  $l_0 = 0.57 \ \mbox{fm}$ which gives a good description of the pion electroproduction data.
The results of the calculation of $T(A)$  for symmetric configuration of 
two pions ($x\sim 0.5$) and $\sigma_{hard}= 5 \ \mbox{mb}$ are presented
in Fig.5 as a function of $p_{\pi}$. One can see from the figure that
the optimal interval for study of the space-time evolution of
the wave packets is $p_{\pi} = 20 \div 40 \ \mbox{GeV}/c$
(we do not consider smaller $p_{\pi}$ since in this case $|t_{min}|$
becomes too large) because $T(A)$ significantly changes
in this $p_{\pi}$ region. 
If CT is observed at high energies for sufficiently asymmetric
configurations, say $x\sim 0.2$, a high precision study of $T(A)$
as a function of $x$ may provide additional tests of the pattern of
space-time evolution of the wave packages. 
In this case the main contribution is given by the expansion of the wave packet forming
a slower pion. In the quantum diffusion model we find that stronger
absorption of a slower pion is partially compensated for by a weaker
absorption of the faster pions resulting in a relatively small
overall change of the transparency in Fig.6.

%%%%%%%%%%%%%%%%%%%%%%%%% figure %%%%%%%%%%%%%%%%%%%%%%%%%
\begin{figure}
\includegraphics[width=0.8\columnwidth]{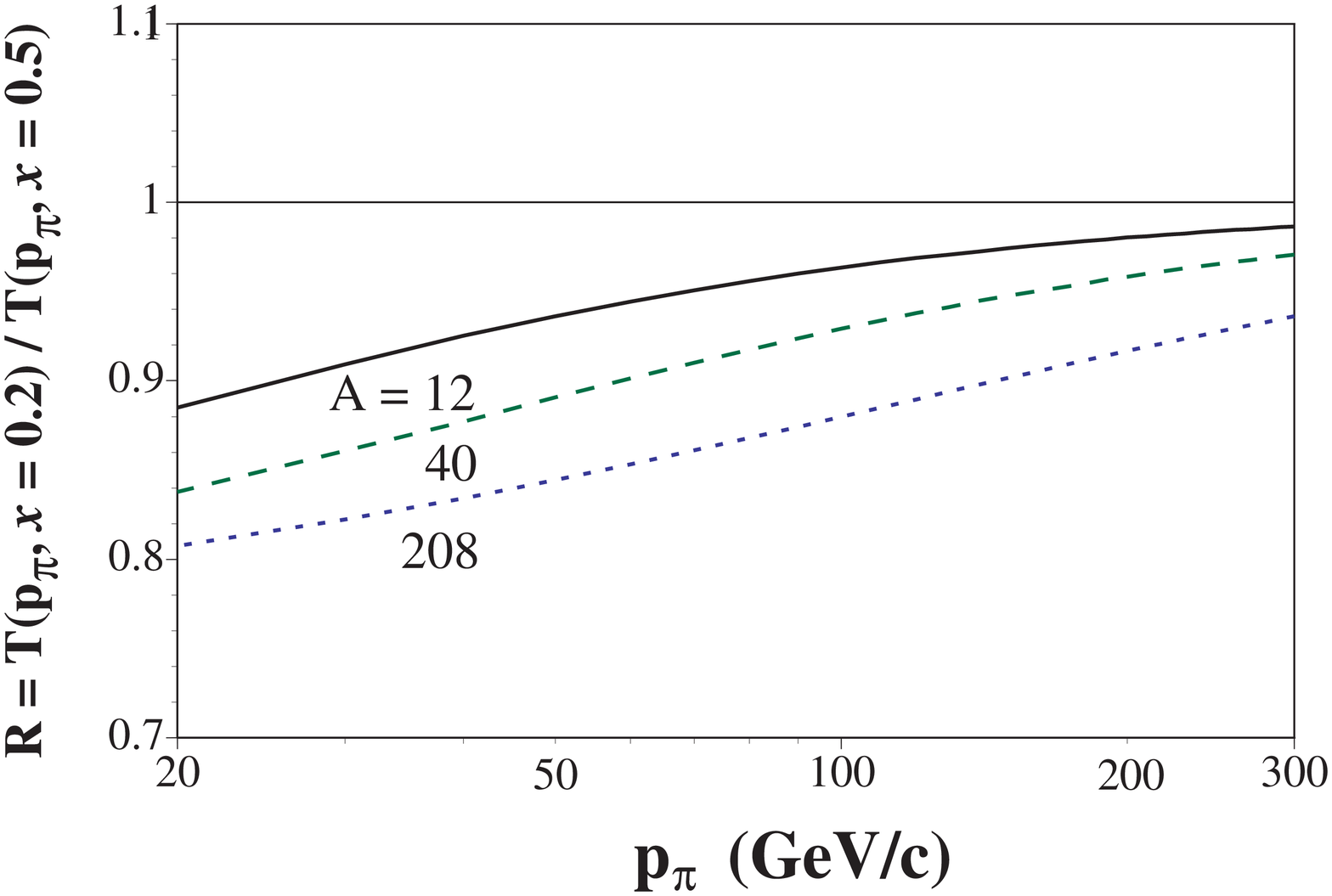}
\vspace{-0.2cm}
\caption{$p_\pi$-dependence of the ratio
$R=T(A)_{x=0.2}/T(A)_{x=0.5}$ for different nuclei. }
\label{diagram:Fig5}
\end{figure}
%%%%%%%%%%%%%%%%%%%%%%%%% figure %%%%%%%%%%%%%%%%%%%%%%%%%

In conclusion,  a  series of measurements for the same configuration of
the two pion system (the same $p_t$ and $M_{\pi \pi}$) corresponding
to the same sizes of the PLCs in the interaction point for a range
of $p_{inc}$ would provide unique information about the space-time
evolution of high energy wave packets.

{\it Other channels}   

Above we focused on the  $\pi^-\pi^+$  state channel. Obviously there
are many other interesting channels. Here we give just few examples: 

$\bullet$  For  an isospin zero target the ratio of $2\to 3$ cross sections for 
production of $\pi^- \pi^- $ and $\pi^- \pi^+ $  in the factorization limit
is equal to the ratio of the $\pi^-\pi^-$ and  $\pi^- \pi^+ $ 
elastic scattering. Based on the observation of the enhancement of
the cross sections of the processes where quark exchange is
allowed \cite{White:1994tj} one could expect that the ratio 
$\sigma_{el}(\pi^-\pi^-)/\sigma_{el}(\pi^- \pi^+)$ is much larger
than one for $-t'/s'\sim 0.5$. Similarly one would be able to measure the ratio of
the $\pi^- \pi^+ $ and $\pi^- \pi^0 $ elastic cross sections.

$\bullet$ One expects to have a significant rate of the process $\pi^++A\to \pi^++K^++A^*$
since the smallness of the  GPDs describing nucleon to hyperon transitions
as compared to the pion-pole dominated GPD is compensated to
some extent by a larger probability for kaon than for pion to be in the PLC which is given by the factor $f_K^2/f_{\pi}^2 \approx 1.4$, 
where $f_\pi$ and $f_K$ are pion and kaon decay constants
\cite{pdg08}.

$\bullet$ One can use  (anti)proton beams to  study the onset of CT
in  $\pi N$ scattering using production of the leading  $N$
and $\pi$ with back to back large transverse momenta.

$\bullet$ One can use high energy proton beams to look for production of
 two back to back protons  with the same   invariant energies as the ones  studied at BNL \cite{Aclander:2004zm}
 to check whether oscillations of the transparency would be present for invariant energies of the proton  pair matching BNL invariant energies.

In summary, we presented a new method for  probing the dynamics
of large angle hadron-hadron scattering using the CT phenomenon which
is free from the limitations imposed by the expansion effects of the PLCs.
It can be applied to a much broader range of two body processes than
the original method including meson-meson scattering ($\pi \pi, \pi K, KK$)
where one expects an earlier onset of the CT regime than for 
meson(baryon)-baryon scattering. One could also look for the  onset of CT in
the meson-baryon ($\pi N, \Lambda \pi, ...$) and baryon-baryon 
($p N, p \Delta,...$) scattering. Studies with beams of  energies
in the 20 $\div$ 200 GeV range appear to be optimal for these purposes
\cite{COMPASS, GSI-FAIR, J-PARC}. If successful, such experiments will open the way to measuring GPDs of a  wide variety of hadrons.

%%%%%%%%%%%%%%%%%%%%%%%%%%%%%%%%%%%%%%%%%%%%%%%%%%%%%%%%%%%%%%%%%%%%%%%%%%%%%%%%

The authors are indebted to A. Carroll for drawing our attention to 
Ref.\cite{Bromberg:1983kv} and to B.~Ketzer and M.~Moinester
for discussion of the COMPASS experiment. Our special thanks are to G.A. Miller for a number of  very helpful suggestions. 
This research has been partially supported by the US DOE Contract
Number DE-FG02-93ER40771 and by the Research Program of Hayama Center
for Advanced Studies of Sokendai.

%%%%%%%%%%%%%%%%%%%%%%%%%%%%%%%%%%%%%%%%%%%%%%%%%%%%%%%%%%%%%%%%%%%%%%%%%%%%%%%%

%%%%%%%%%%%%%%%%%%%%%%%%%%%%%%%%%%%%%%%%%%%%%%%%%%%%%%%%%%%%%%%%%%%%%%%%%%%%%%%%

\end{document}